\documentclass[prl,superscriptaddress,showpacs,twocolumn,10pt]{revtex4}
\usepackage{amssymb}
\usepackage{graphicx}

\def\>{\rangle}

\begin{document}
\newtheorem{corollary}{Corollary}
\newtheorem{definition}{Definition}
\newtheorem{example}{Example}
\newtheorem{lemma}{Lemma}
\newtheorem{proposition}{Proposition}
\newtheorem{statement}{Statement}
\newtheorem{theorem}{Theorem}
\newtheorem{property}{Property}
\newtheorem{fact}{Fact}
\newtheorem{conjecture}{Conjecture}

\newcommand{\bra}[1]{\langle #1|}
\newcommand{\ket}[1]{|#1\rangle}
\newcommand{\braket}[3]{\langle #1|#2|#3\rangle}
\newcommand{\ip}[2]{\langle #1|#2\rangle}
\newcommand{\op}[2]{|#1\rangle \langle #2|}

\newcommand{\tr}{{\rm tr}}
\newcommand {\E } {{\mathcal{E}}}
\newcommand {\F } {{\mathcal{F}}}
\newcommand {\M} {{\mathcal{M}}}
\newcommand {\f } {\tilde{F}}
\newcommand {\rr } {\tilde{r}}
\newcommand {\R } {{\mathcal{R}}}
\newcommand {\I } {{\mathcal{I}}}
\newcommand{\mod}{{\rm\  mod\ }}
\renewcommand{\b}{\mathcal{B}}
\newcommand{\h}{\mathcal{H}}
\newcommand{\T}{\mathcal{T}}

\title{The Perfect Distinguishability of Quantum Operations}
\author{Runyao Duan}
\email{Runyao.Duan@uts.edu.au}

\author{Yuan Feng}

\author{Mingsheng Ying}

\affiliation{State Key Laboratory of Intelligent Technology and
Systems,\\ Tsinghua National Laboratory for Information Science and
Technology,\\ Department of Computer Science and Technology,\\
Tsinghua
University, Beijing 100084, China\\
{and}
\\
Center for Quantum Computation and Intelligent Systems (QCIS),
Faculty of Engineering and Information Technology, University of
Technology, Sydney, NSW 2007, Australia}

\date{\today}

\begin{abstract}
We provide a feasible necessary and sufficient condition for when an
unknown quantum operation (quantum device) secretely selected from a
set of known quantum operations can be identified perfectly within a
finite number of queries, and thus complete the characterization of
the perfect distinguishability of quantum operations. We further
design an optimal protocol which can achieve the perfect
discrimination between two quantum operations by a minimal number of
queries. Interestingly, employing the techniques from the theory of
$q$-numerical range we find that an optimal perfect discrimination
between two isometries is always achievable without using auxiliary
systems or entanglement.
\end{abstract}


\maketitle

\textit{1. Introduction:} One of the fundamental features of quantum
mechanics is that it is impossible to distinguish between two
nonorthogonal states perfectly, even when arbitrarily large but
finite copies of states are available. A recent highlight of this
fact is the identification of the quantum Chernoff
bound~\cite{ACMB+07}. In view of this, it is clear that the perfect
distinguishability of quantum states is completely characterized by
the orthogonality.

A problem closely related to quantum state discrimination is
consider the distinguishability of quantum operations (or
intuitively quantum devices), which formalize all physically
realizable operations in quantum mechanics. The basic problem can be
described as follows. Assume we are given an {\it unknown quantum
device} which belongs to one of two known quantum operations. Our
purpose is to figure out the identity of this device by  a finite
number of queries together with any other allowable physical
operations. This problem has received great interest in recent years
and a number of results have been reported (See Refs.
\cite{AKN97,CP00, AC01,SAC05,DFY07,WD08,JFDY06} for a partial list
of these works). It has been shown that distinguishing quantum
operations has many interesting properties that are similar to that
of quantum state discrimination if the device is probed only once
\cite{AKN97,CP00,AC01,SAC05}. On the other hand, quantum devices are
very different from quantum states in the following three aspects.
First, a quantum device is reusable. Second, the input state of
quantum device can be chosen freely, and thus can be entangled with
an auxiliary system or between different uses. Third, perhaps most
importantly, a quantum device can be used in many essentially
different ways such as in parallel, in sequential, or in any other
scheme allowed by quantum mechanics while the optimal way to
manipulate many copies of quantum states is uniquely in parallel.
Due to these differences, it is quite difficult to identify the
behavior of quantum operations when multiple queries are used. In
particular, it is unclear when two quantum operations are perfectly
distinguishable within a finite number of queries.

Several works have been devoted to the perfect distinguishability of
special quantum operations including unitary
operations\cite{CP00,AC01,DFY07,WD08} and projective measurements
\cite{JFDY06}. Most notably, any two different unitary operations
can be perfectly distinguishable by inputting an entangled state and
applying the unknown unitary in parallel \cite{AC01}. Such a perfect
discrimination can also be achieved by applying the unitary
operations on a single system sequentially, and entanglement or
joint quantum operations are not necessary \cite{DFY07}.
Interestingly, projective measurements also enjoy this kind of
perfect distinguishability \cite{JFDY06}. Very recently experimental
results concerning with the perfect discrimination of unitary
operations and measurements have been reported \cite{LRB09}. All
these progresses  indicate that the notion of perfect
distinguishability of general quantum operations would be much more
complicated than that of quantum states. The minimum-error or
unambiguous discrimination strategies for quantum states cannot be
simply applied to quantum operations as they cannot fully reflect
the fact that many quantum operations are essentially perfectly
distinguishable in multi-use scenario although a perfect
discrimination cannot be achieved by one single use.

The purpose of this Letter is to provide a complete characterization
of the perfect distinguishability of quantum operations (See Theorem
\ref{qo} below). We show that two simple properties are necessary
and sufficient for the perfect discrimination between two quantum
operations within a finite number of queries. The first property
says that two quantum operations that are perfectly distinguishable
should produce two quantum states with non-overlapping supports upon
some common input state, which may entangled with an auxiliary
system. The second property states that any such two quantum
operations are capable of transforming some two non-orthogonal pure
states, which are provided to the quantum operations as their
respective inputs, into orthogonal states. These two properties
reveal the key feature of the perfect distinguishability of quantum
operations and thus provide new insight into this problem. It is
also worth noting that both of these properties can be rephrased
into analytical forms in terms of the Kraus operators of quantum
operations to distinguish and can be verified quite efficiently. As
a potential application, we show that the classical data hiding is
possible by encoding the data into quantum devices instead of
quantum states \cite{TDL01}.

Furthermore, with the assistance of a mathematical notion of the
maximal fidelity between quantum states, we can provide an optimal
protocol which can distinguish two quantum operations with a minimal
number of queries. This number can be efficiently determined using
numerical iteration techniques. We further show that for
distinguishing between two isometries (generalization of unitary
operations), an optimal discrimination always can be achieved
without auxiliary systems or entanglement by employing some results
from the theory of $q$-numerical range. This generalizes our
previous work on unitary operations \cite{DFY07}.

\textit{2. Conditions for the perfect discrimination between quantum
operations} Consider a $d$-dimensional Hilbert space $\h_d$. The set
of linear operators on $\h_d$ is denoted by $\b(\h_d)$. A general
quantum state $\rho$ on $\h_d$ is given by a positive operator in
$\b(\h_d)$ with trace one. A pure state $\ket{\psi}$ is a unit
vector in $\h_d$. For simplicity, we will use $\psi$ to denote the
density operator form $\op{\psi}{\psi}$ of $\ket{\psi}$. Let $\rho$
be with the spectral decomposition $\rho=\sum_{k=1}^d
p_k\op{\psi_k}{\psi_k}$. The support of $\rho$ is given by ${\rm
supp}(\rho)={\rm span}\{\ket{\psi_k}:p_k>0\}$. A quantum operation
$\E$ from $\b(\h_d)$ to $\b(\h_{d'})$ is a {\it trace-preserving
completely positive map} with the the form $\E(\rho)=\sum_{i=1}^m
E_i\rho E_i^\dagger$, where $\{E_i\}_{i=1\cdots m}$ are the Kraus
operators of $\E$ satisfying $\sum_{i}E_i^\dagger E_i=I_d$. Quantum
operations formalize all physically realizable operations allowed by
quantum mechanics, including unitary operations, quantum
measurements, and quantum channels. In particular, a quantum
measurement $\M$ with measurement operators $\{E_1,\cdots, E_m\}$ is
a special quantum operation with Kraus operations $\{E_k\otimes
\ket{k}:k=1\cdots m\}$, where $\{\ket{k}\}$ is a classical system
with $m$ distinguishable states. To emphasize the importance of the
order among the measurement operators, a quantum measurement $\M$
can be represented as an $m$-tuple of matrices, say $(E_1,\cdots,
E_m)$.

Two density operators $\rho_0$ and $\rho_1$ are said to be disjoint
if ${\rm supp}(\rho_0)\cap {\rm supp}(\rho_1)=\{0\}$. Let us now
introduce a notion to quantitatively describe the disjointness
between two quantum states, which can a treated as a special inner
product between two mixed states (actually two subspaces).

\begin{definition}\label{max-fidelity}\upshape
The \textit{maximal fidelity} between two quantum states $\rho_0$
and $\rho_1$ is defined as follows:
$$
\f(\rho_0,\rho_1)=\max\{|\ip{\psi_0}{\psi_1}|: \ket{\psi_k}\in{\rm
supp}(\rho_k),k=0,1\}
$$
\end{definition}
It follows from the definition that $0\leq \f(\rho_0,\rho_1)\leq 1$.
$\f$ is vanishing iff $\rho_0$ and $\rho_1$ are orthogonal, and
attains $1$ iff $\rho_0$ and $\rho_1$ are not disjoint. The name
``maximal fidelity" comes from the following simple connection to
the ordinary fidelity:
$$\f(\rho_0,\rho_1)=\max\{F(\rho'_0,\rho_1'): {\rm supp}(\rho_k')\subseteq {\rm supp}(\rho_k)\},$$
where $F(\rho_0,\rho_1)=\tr\sqrt{\rho_0^{1/2}\rho_1\rho_0^{1/2}}$.
Due to the above connection, the maximal fidelity $\f$ enjoys some
similar properties as $F$. For instance, for pure states both of
them coincide with the ordinary inner product, and the maximal
fidelity is also multiplicative according to tensor product. The
most important property of the maximal fidelity is the following
operational interpretation. Note that a similar operational
interpretation of $F(\rho_0,\rho_1)$ has been found in \cite{DN02}.
The technical proof is put in the appendix.

\begin{lemma}\label{key}\upshape
For two pairs of quantum states  $\{\rho_0,\rho_1\}$ and
$\{\ket{\psi_0},\ket{\psi_1}\}$, there is a quantum operation $\T$
such $\T(\rho_k)=\psi_k$ for $k=0,1$ iff $\f(\rho_0,\rho_1)\leq
\f({\psi_0},{\psi_1})=|\ip{\psi_0}{\psi_1}|$. Thus we have
\begin{equation}\label{max-fid-meaning}
\f(\rho_0,\rho_1)=\min\{|\ip{\psi_0}{\psi_1}|: \exists \T,
\T(\rho_k)=\psi_k\}.
\end{equation}
\end{lemma}

It is straightforward to define two quantum operations are disjoint.
Formally, we have the following
\begin{definition}\label{qo-disjoint}\upshape
$\E_0$ and $\E_1$ are said to be {\it (unassisted) disjoint} if
there is an input state $\ket{\psi}\in\h_d$ such that $\E_0(\psi)$
and $\E_1(\psi)$ are disjoint. $\E_0$ and $\E_1$ are said to be {\it
entanglement-assisted disjoint} if there is an input state
$\ket{\psi}^{RQ}$ such that $(\I^R\otimes \E_0^Q)(\psi^{RQ})$ and
$(\I^R\otimes \E_1^Q)(\psi^{RQ})$ are disjoint, where $R$ and $Q$
denote auxiliary and principal systems respectively, and $\I^R$ is
the identity operation on $R$.
\end{definition}

One can easily verify that the dimension of $R$ in the above
definition can be assumed to be the same as $Q$ and larger dimension
cannot make any difference.

There is an efficient procedure to determine whether two quantum
operations $\E_0$ and $\E_1$ are entanglement-assisted disjoint.
Suppose that $\mathcal{S}_k={\rm span}\{E_{0i}\}_{i=1\cdots n_k}$,
$k=0,1$. If $\mathcal{S}_0\cap\mathcal{S}_1=\{0\}$ then $\E_0$ and
$\E_1$ are entanglement-assisted disjoint and the input state can be
chosen as
$\ket{\alpha}^{RQ}=1/\sqrt{d}\sum_{k=1}^d\ket{k}^R\ket{k}^Q$.
Otherwise, select an arbitrary basis $\{D_i\}_{i=1\cdots p}$ for
$\mathcal{S}_0\cap\mathcal{S}_1$, and construct an operator
$X=\sum_{k=1}^p D_k^\dagger D_k$. Let $P_1$ be the projector onto
${\rm supp}(X)$, and consider two new channels $\E_0'$ and $\E_1'$
with respective Kraus operators $\{E_{0i}P_1^\perp\}$ and
$\{E_{1j}P_1^\perp\}$, where $P_1^\perp=I_d-P_1$. The original
problem is now reduced to decide whether $\E_0'$ and $\E_1'$ are
entanglement-assisted disjoint, and a projector $P_2\leq P_1^\perp$
can be similarly constructed. Repeat this process $n\leq d$ times we
can efficiently construct a sequence of {\it mutual orthogonal}
projectors $P_1,\cdots, P_n$ such that $P_n=0$ and $P_{i}\neq 0$ for
any $i<n$. Let $P=I_d-\sum_{i=1}^{n-1}P_i$. Then  $\E_0$ and $\E_1$
are entanglement-assisted disjoint iff $P\neq 0$. If satisfied,
$\ket{\psi}=(I\otimes P)\ket{\alpha}$ is an eligible input state.

We are now ready to present a complete characterization of the
perfect distinguishability of quantum operations.

\begin{theorem}\label{qo}\upshape
Let $\E_0$ and $\E_1$ be two quantum operations from $\b(\h_d)$ to
$\b(\h_{d'})$ with Kraus operators $\{E_{0i}:i=1\cdots n_0\}$ and
$\{E_{1j}:{j=1\cdots n_1}\}$, respectively. Then $\E_0$ and $\E_1$
are perfectly distinguishable by a finite number of uses iff i)
$\E_0$ and $\E_1$ are entanglement-assisted disjoint, and ii)
$I_d\not\in {\rm span}\{E_{0i}^\dagger E_{1j}\}$.
\end{theorem}

{\bf Proof.} Let us first show show that the conditions i) and ii)
are necessary. Suppose $\E_0$ and $\E_1$ are perfectly
distinguishable within $N$ uses, and assume $N$ is minimal. We claim
that there is an input state $\ket{\psi}^{RQ}$ such that
$(\I^R\otimes \E_0^Q)(\psi)$ and $(\I^R\otimes \E_1^Q)(\psi)$ are
disjoint. By contradiction, assume that for any choice of
$\ket{\psi}^{RQ}$, $(\I^R\otimes \E_0^Q)(\psi)$ and $(\I^R\otimes
\E_1^Q)(\psi)$ are not disjoint. Then we can find a state
$\ket{\psi'}^{RQ}$ that lies in both supports. Thus in the next
$N-1$ uses we must be able to distinguish between $\E_0$ and $\E_1$
by inputting $\ket{\psi'}^{RQ}$. That means $(N-1)$ uses are
sufficient to distinguish between $\E_0$ and $\E_1$ by inputting
$\ket{\psi'}^{RQ}$. This contradicts the minimality of $N$. Hence
$\E_0$ and $\E_1$ must be entanglement-assisted disjoint.

To show the necessity of ii), let's consider the last use of the
unknown quantum operation. Assume that the input states
corresponding to $\E_0$ and $\E_1$ are $\rho_0$ and $\rho_1$,
respectively. Both $\rho_0$ and $\rho_1$ are the output states of
previous $(N-1)$ uses and may be mixed states. As the last use must
distinguish between $\E_0$ and $\E_1$ but all previous $(N-1)$ uses
cannot, we have $(\I\otimes \E_0)(\rho_0)\perp
(\I\otimes\E_1)(\rho_1),~\rho_0\not \perp\rho_1.$ Thus there must be
two states $\ket{\psi_0}=(\I^R\otimes A_0^Q)\ket{\alpha}^{RQ}$ and
$\ket{\psi_1}=(I^R\otimes A_1^Q)\ket{\alpha}^{RQ}$ from the supports
of $\rho_0$ and $\rho_1$, respectively, such that
$$\tr((\I\otimes \E_0)(\psi_0)(\I\otimes\E_1)(\psi_1))=0,~{\rm and}~\ip{\psi_0}{\psi_1}\neq 0.$$
Substituting the Kraus operators of $\E_0$ and $\E_1$ into the above
equation, we have
$$\tr(E_{0i}^\dagger E_{1j} A_1A_0^\dagger)=0,\forall i,j,~{\rm and}~\tr(A_1A_0^\dagger)\neq 0.$$
That is the same as $I_d\not\in {\rm span}\{E_{0i}^\dagger
E_{1j}\}$.

Sufficiency part can be proven by constructing a protocol to
distinguish between $\E_0$ and $\E_1$ as follows:

Step 1. Calculate a pair of pure states $\ket{\psi_0}^{RQ}$ and
$\ket{\psi_1}^{RQ}$ such that $\ip{\psi_0}{\psi_1}\neq 0$ and
$(\I\otimes \E_0)(\psi_0)\perp (\I\otimes \E_1)(\psi_1)$. This
always can be done due to condition ii).  More precisely, we can
first choose a matrix $M\in {\rm span}^\perp\{E_{0i}^\dagger
E_{1j}\}$ such that $\tr(M)\neq 0$. Then let $A_0=I_d/d$ and
$A_1=M/\sqrt{\tr(M^\dagger M)}$, and construct
$\ket{\psi_0}=(I\otimes A_0)\ket{\alpha}$ and
$\ket{\psi_1}=(I\otimes A_1)\ket{\alpha}$. Clearly, such
$\ket{\psi_0}$ and $\ket{\psi_1}$ satisfy our requirements.

Step 2. Choose a state $\ket{\phi}^{RQ}$ such that
$\rho_0=(\I\otimes\E_0)(\phi)$ and $\rho_1=(\I\otimes \E_1)(\phi)$
are disjoint. This can be done due to condition i). Furthermore,
such a state $\ket{\phi}$ can be efficiently determined by the
procedure below Definition \ref{qo-disjoint}.

Prepare $N$ copies of $\ket{\phi}^{RQ}$ and apply the unknown
quantum operations $N$ times to each copy in parallel. Then we are
left with a quantum state belonging to $\{\rho_0^{\otimes
N},\rho_1^{\otimes N}\}$. Choose $N$ to be the minimal integer such
that $\f(\rho_0^{\otimes N},\rho_1^{\otimes
N})=\f(\rho_0,\rho_1)^N\leq |\ip{\psi_0}{\psi_1}|$. We can choose
$N=\lceil\ln |\ip{\psi_0}{\psi_1}|/\ln \f(\rho_0,\rho_1)\rceil$.
(Note that $\rho_0$ and $\rho_1$ are disjoint, thus
$\f(\rho_0,\rho_1)<1$)

Step 3. Transform $(\rho_0^{\otimes N},\rho_1^{\otimes N})$ into
$(\ket{\psi_0},\ket{\psi_1})$ by some quantum operation $\T$, which
can be done due to our choice of $N$ and Lemma \ref{key}. Then
applying the unknown quantum operation to
$(\ket{\psi_0},\ket{\psi_1})$ one more time will yield two
orthogonal states, which will allow us to perfectly identify the
unknown quantum operation with $N+1$ queries.\hfill $\blacksquare$

It is worth noting that by the standard arguments in
\cite{CP00,AC01,DFY07} Theorem \ref{qo} can be directly extended
into the case where the number of quantum operations to be
distinguished is more than two.

Applying Theorem \ref{qo} to specific quantum operations, we can
directly obtain many interesting results on the perfect
distinguishability of quantum operations, which include previous
results regarding the perfect discrimination of unitary operations
\cite{AC01,DFY07} and projective measurements \cite{JFDY06}. As a
new example, let us consider the discrimination between an isometry
and a quantum operation.  Note that an isometry is a linear operator
$U$ from $\h_d$ to $\h_{d'}$ such that $U^\dagger U=I_d$. One can
easily verify that an isometry $U$ and a general quantum operation
$\E=\sum_{k=1}^n E_k\cdot E_k^\dagger$ are perfectly distinguishable
iff $I_d\not\in {\rm span}\{U^\dagger E_k:k=1\cdots n\}$. In
particular, a unitary $U$ and $\E$ are perfectly distinguishable if
and only if $U\not\in {\rm span}\{E_k\}$, i.e., $U$ cannot be a
Kraus operator of $\E$. In all these cases only condition ii) is
involved as it is stronger than condition i).  This is not true in
general. For instance, one can write down the conditions for the
perfect discrimination between two general quantum measurements,
which have a simpler form in terms of measurement operators. In this
case conditions i) and ii) are independent and none of them can be
removed.

A potential application of Theorem \ref{qo} is to design the
following classical data hiding protocol: A boss, say Charlie,
encodes a secret task (described as a secret bit $b$) into two pairs
of quantum operations (devices) $(\E_b,\E_b')_{b=0,1}$, and
allocates $\E_b$ and $\E_b'$ to two distant employees, Alice and
Bob, respectively. Alice and Bob are not allowed to individually
exactly recover $b$ while Charlie can reveal the bit at any time by
supplying entanglement or asking them to move together. This kind of
protocol has been shown to be possible if Charlie encodes the bit
using two orthogonal bipartite mixed states $\rho_0^{AB}$ and
$\rho_1^{AB}$ that are locally indistinguishable \cite{TDL01}. The
new feature of hiding classical data using quantum devices instead
of states is that the identified device can be reused in the future
information processing tasks.

Applying Theorem \ref{qo}, we can easily construct these kind of
instances by imposing that $\{\E_0,\E_1\}$ satisfies only condition
i) while $\{\E_0',\E_1'\}$ satisfies only condition ii). An explicit
example is as follows: $\E_0$ and $\E_1$ are quantum operations that
prepare quantum states
$\ket{\psi_0}=(\ket{0}+\sqrt{2}\ket{1})/\sqrt{3}$ and
$\ket{\psi_1}=(\ket{0}-\sqrt{2}\ket{1})/\sqrt{3}$, respectively;
while $\E_0'=(\op{0}{0}+1/\sqrt{2}\op{1}{1},1/\sqrt{2}\op{1}{1},0)$
and $\E_0'=(\op{0}{0}+1/\sqrt{2}\op{1}{1},0,1/\sqrt{2}\op{1}{1})$
are two one-qubit measurements. One can easily verify that $\E_0'$
and $\E_1'$ are perfectly distinguishable upon the respective input
states $\ket{\psi_0}$ and $\ket{\psi_1}$ as
$\tr{((\op{0}{0}+1/2\op{1}{1})\op{\psi_0}{\psi_1})}=0$.

\textit{3. An optimal protocol for the perfect discrimination
between two quantum operations} The discrimination protocol we
presented in Theorem \ref{qo} is not optimal in general. We shall
now describe an optimal one. We need a notion of $q$-maximal
fidelity, which is naturally induced from the maximal fidelity
between quantum states, to quantitatively describe the disjointness
between quantum operations.

\begin{definition}\label{qo-fidelity}\upshape
For quantum operations $\E_0$ and $\E_1$, and $0\leq q\leq 1$, the
{\it $q$-maximal fidelity} is defined as follows:
$$
\f_q(\E_0,\E_1)=\min\{\f(\E_0(\psi_0),\E_1(\psi_1)):
\ip{\psi_0}{\psi_1}=q\}.
$$
The {\it entanglement-assisted $q$-maximal fidelity} is defined as
follows
$$\f^{ea}_q(\E_0,\E_1)=\f_q(\I^R\otimes \E_0^Q,\I^R\otimes \E_1^Q),$$
where $R$ is an auxiliary system with the same dimension as $Q$
(larger cannot make difference). When $q=1$, $\f_1(\E_0,\E_1)$ and
$\f_1^{ea}(\E_0,\E_1)$ are said to be the maximal fidelity and the
entanglement-assisted maximal fidelity between $\E_0$ and $\E_1$,
respectively.
\end{definition}
Here we should point out that $\psi_0$ and $\psi_1$ in the above
definition can be replaced by any $\rho_0$ and $\rho_1$ such that
$\f(\rho_0,\rho_1)=q$. However, in virtue of Lemma \ref{key} we can
verify that it is sufficient to consider pure states only.

The notion of $\f_q^{ea}(\E_0,\E_1)$ plays a crucial role in
designing the optimal perfect discrimination protocol of quantum
operations, which is mainly due to the following desirable property:
\begin{equation}\label{strong-mono}
\f_q^{ea}(\E_0,\E_1)\leq \frac{q}{q'}\f_{q'}^{ea}(\E_0,\E_1),0\leq
q<q'\leq 1.
\end{equation}
This property can be understood as``more separable states will yield
more separable output states." It is true simply due to the fact
that by appending an auxiliary qubit we can divide the input states
for $f_q^{(ea)}$ into two parts: a pair of qubit states with inner
product $q/q'$ and a pair of optimal input states for $f_{q'}^{ea}$.

Let us start to describe an optimal perfect discrimination protocol
between $\E_0$ and $\E_1$. Let $N_{\min}$ be the minimal number of
uses of the unknown quantum operation required to perfectly
distinguish between $\E_0$ and $\E_1$, and let $\{q_k\}$ be a
sequence of $q$-maximal fidelities recursively defined as follows:
$$q_0=1, q_k=\f_{q_{k-1}}^{ea}(\E_0,\E_1),
k\geq 1.$$ Notice that $q_1=\f^{(ea)}_1(\E_0,\E_1)$ is just the
entanglement-assisted maximal fidelity between $\E_0$ and $\E_1$.
Let us further introduce $q_{\max}$ as follows:
$$q_{\max}=\{q:\f_{q}^{(ea)}(\E_0,\E_1)=0\}.$$ Then the following theorem shows
that $N_{\min}$ is completely determined by the sequence of
$\{q_k\}$ and $q_{\max}$ (indirectly).

\begin{theorem}\upshape
Let $\mathcal{N}^{(k)}$ represent an arbitrary quantum
discrimination network containing $k$ uses of the unknown quantum
operation from $\{\E_0,\E_1\}$. Then
$$q_k\leq \f_1^{ea}(\mathcal{N}^{(k)}(\E_0),\mathcal{N}^{(k)}(\E_1)).$$
In other words, $q_k$ is the optimal maximal fidelity one can
achieve by $k$ uses of the unknown quantum operation from
$\{\E_0,\E_1\}$ and with the same input. Furthermore,
$N_{\min}=\min\{k: q_k=0,k\geq 1\}=\min\{k: q_{k-1}\leq q_{\max}\}$,
and $q_k=0$ for any $k>{N_{\min}}$.
\end{theorem}

{\bf Proof.} By mathematical induction. By definition $q_1$ is the
optimal maximal fidelity one can achieve by a single use. Assume
that $q_k$ is optimal by $k$ uses of the unknown quantum operation.
Consider any quantum discrimination network $\mathcal{N}^{(k+1)}$
containing $k+1$ uses of the unknown quantum operation. By induction
assumption, We have $q_k'=\f(\rho^{(k)}_0, \rho^{(k)}_1)\geq q_k$,
where $\rho^{(k)}_0$ and $\rho^{(k)}_1$ are the output states of
$\mathcal{N}^{(k+1)}$ except the last use of the unknown quantum
operation. Clearly, $\rho^{(k)}_0$ and $\rho^{(k)}_1$ are the output
states of a quantum discrimination network containing $k$ uses of
the unknown quantum operation, and also the input states for the
last use of the unknown quantum operation in $\mathcal{N}^{(k+1)}$.
Let $\rho_0^{(k+1)}$ and $\rho_1^{(k+1)}$ be the final output states
of $\mathcal{N}^{(k+1)}$.  By Eq. (\ref{strong-mono}), we have
$$
\f(\rho_0^{(k+1)},\rho_1^{(k+1)})\geq \f^{ea}_{q_k'}(\E_0,\E_1)\geq
\frac{q_k'}{q_k}\f^{ea}_{q_k}(\E_0,\E_1)\geq q_{k+1},
$$
where we have employed the assumption $q_k'\geq q_k$ and the
definition of $q_{k+1}$. The expression of $N_{\min}$ follows
immediately. \hfill $\blacksquare$

It is clear from the above proof that $q_1$ and $q_{\max}$ are
responsible for the perfect discrimination between  $\E_0$ and
$\E_1$. More precisely, $\E_0$ and $\E_1$ are perfectly
distinguishable iff $q_1<1$ and $q_{\max}>0$, which is based on the
following two simple observations: 1) $q_1=1$ implies $q_k=1$ for
any $k\geq 1$; or 2) $q_{\max}=0$ implies $q_k>0$ for any $k\geq 1$.
One can also readily verify that $q_1<1$ and $q_{\max}>0$ correspond
to conditions i) and ii) in Theorem \ref{qo}, respectively. As a
consequence, we can obtain an upper bound of $N_{\min}$ in terms of
$q_1$ and $q_{\max}$, say $N_{\min}\leq \lceil\ln q_{\max}/\ln
q_1\rceil$. Note here $q_1=0$ implies $N_{\min}=1$.

The sequence of $\{q_k\}$ and $q_{\max}$  can be calculated with
arbitrary high precision using numerical iteration techniques as it
is evident that $F_q^{(ea)}(\E_0,\E_1)$ can be formulated into an
optimization problem on a compact set. Hence we can estimate
$N_{\min}$ for any two quantum operations $\E_0$ and $\E_1$
according to the above theorem. In many practical applications, a
simple protocol like the one in Theorem \ref{qo} would be
sufficient.

\textit{4. $q$-numerical range and the perfect distinguishability of
isometries} For general $\E_0$ and $\E_1$, it is normally very
difficult to calculate the optimal fidelity sequence of $\{q_k\}$.
Interestingly, if both $\E_0$ and $\E_1$ are isometries, the
calculation becomes quite tractable. For isometries $U_0$ and $U_1$,
we have
$$\f_{q}(U_0,U_1)=\rr_q(A)=\min\{|z|:z\in W_q(A)\},$$
where $A=U_0^\dagger U_1$ \cite{A-isometry} and
$W_q(A)=\{\braket{\psi_0}{A}{\psi_1}:\ip{\psi_0}{\psi_1}=q\}.$
Similarly, $\f_{q}^{ea}(U_0,U_1)=\rr_q(I_d\otimes A).$ For $0\leq
q\leq 1$, $W_q(A)$ is said to be the $q$-numerical range of $A$ with
$\rr(A)$ the inner radius. When $q=1$, $W(A)=W_1(A)$ is the classic
numerical range of $A$. The theory of numerical range and its
various generalizations including $q$-numerical range are an active
and vast topic in linear algebra \cite{HJ91}. It has been recognized
recently that these notions are quite useful in studying the local
discrimination of unitary operations \cite{DFY08}. A somewhat
surprising fact is that the optimal perfect discrimination of
isometries can be achieved without auxiliary systems or
entanglement.

\begin{theorem}\label{opt-isometry}\upshape
For any isometries $U_0$ and $U_1$, and $0\leq q\leq 1$,
$\f_q^{ea}(U_0,U_1)=\f_q(U_0,U_1)$.
\end{theorem}
Previously we have shown the same result for unitary operations
\cite{DFY07}. We can derive the above result from an interesting
result about the $q$-numerical range, say $W_q(I_d\otimes A)=W_q(A)$
for any linear operator $A$ and $0\leq q\leq 1$. The equality for
the case of $q=1$ follows directly from the convexity of $W(A)$. For
the general case we cannot find any existing reference to this
important result and thus we provide a proof in the appendix.

There is no explicit expression for the $q$-inner radius $\rr(A)$ of
a general linear operator $A$. Hence it is generally impossible to
obtain the analytical formula of $N_{\min}(U_0,U_1)$. Fortunately,
it was known that $W_q(A)$ is a convex compact set for any linear
operator $A$ and $0\leq q\leq 1$, and efficient characterization of
the boundary of $W_q(A)$ has been obtained \cite{Tsi84}. As a
consequence, it is quite feasible to compute $\rr_q(A)$, and then
determine the exact value of $N_{\min}$. It is also possible to
obtain analytical results when $A$ belongs to normal or $2\times 2$
matrices as efficient characterization of the $q$-numerical range
has been found. In particular, the case that $A$ is unitary has been
completely solved \cite{DFY07}. For the case that $A$ is positive
definite, however, any parallel protocol cannot distinguish between
$U_0$ and $U_1$, even assisted with arbitrary large amount of
entanglement. In sharp contrast, we know from Theorem
\ref{opt-isometry} that there is a sequential protocol that can
achieve an optimal perfect discrimination. Furthermore, in this case
$W_q(A)$ is an elliptical disk with eccentricity $q$, and foci
$q\lambda_0$ and $q\lambda_1$, where $\lambda_0$ and $\lambda_1$ are
the maximum and minimum of eigenvalues of $A$ \cite{Tsi84}. Using
this fact one can derive the following analytical formula:
$$N_{\min}(U_0,U_1)=\lceil\frac{\ln 2+\ln (1-\lambda_1)-\ln(\lambda_1)}{\ln 2-\ln (\lambda_0+\lambda_1)}\rceil.$$

\textit{5. Discussions} It would be highly desirable to identify the
quantum Chernoff bound for quantum operations that are not perfectly
distinguishable. Perhaps the first step to this problem is to
identify the (asymptotically) optimal minimum-error discrimination
strategy for quantum operations using the distance measure induced
by diamond norm instead of the maximal fidelity. Many of our
techniques can be generalized to multipartite setting, where distant
parties share an unknown quantum operation and they are only allowed
to perform arbitrary Local Operations and Communicate with each
other Classically (LOCC). In a previous work we have shown that the
perfect distinguishability of unitary operations is preserved under
LOCC \cite{DFY08}, benefitting from the local distinguishability of
two orthogonal multipartite pure states \cite{WSHV00}. With some
additional efforts we can generalize Theorem \ref{qo} to a wider
class of multipartite quantum operations, including all isometries
and almost all quantum measurements. Unfortunately, the condition
for the perfect distinguishability of general multipartite quantum
operations remains unknown as it is still unknown when two general
orthogonal mixed states can be locally distinguishable. We will
continue to study these issues.

This work was partially supported by the National Natural Science
Foundation of China (Grant Nos. 60702080, 60736011), the Hi-Tech
Research and Development Program of China (863 project) (Grant No.
2006AA01Z102), and the FANEDD under Grant No. 200755.

\section*{Appendix}
{\bf Proof of Lemma \ref{key}.} The result is true when both
$\rho_0$ and $\rho_1$  are pure states \cite{Uhl85}. We only need to
focus on general case.

Necessity: By definition, there are two pure states $\ket{a_0}\in
{\rm supp}(\rho_0)$ and $\ket{a_1}\in {\rm supp}(\rho_1)$ such that
$|\ip{a_0}{a_1}|=\f(\rho_0, \rho_1)$. Thus we have $\T(a_0)=\psi_0$
and $\T(a_1)=\psi_1$. It follows immediately from Ref. \cite{Uhl85}
that $\f(\rho_0,\rho_1)=|\ip{a_0}{a_1}|\leq |\ip{\psi_0}{\psi_1}|$.

Sufficiency: Assume that $\f(\rho_0,\rho_1)\leq
|\ip{\psi_0}{\psi_1}|$. We will construct a quantum operation $\T$
such that $\T(\rho_k)=\psi_k$ for $k=0,1$. We exclude the trivial
cases and assume $0<\f(\rho_0,\rho_1)<1$. Let $P$ and $Q$ be the
projectors onto ${\rm supp}(\rho_0)$ and ${\rm supp}(\rho_1)$,
respectively. By assumption we know that $PQ\neq 0$. Applying the
singular-valued decomposition theorem to $PQ$, we have
$$PQ=\sum_{k=1}^r \lambda_k \op{\psi_0^{(k)}}{\psi_1^{(k)}},$$
where $\lambda_k>0$ for each $k=1\cdots r$ and $r$ is the rank of
$PQ$. By the above equation and the fact that both $P$ and $Q$ are
projectors, we know that $\{\ket{\psi_0^{(k)}}\}$ is a set of
orthonormal states from ${\rm supp}(P)$. Similarly
$\{\psi_1^{(k)}\}$ is a set of orthonormal states from ${\rm
supp}(Q)$. Further more, we have
$\ip{\psi_0^{(i)}}{\psi_1^{(j)}}=\delta_{ij}\lambda_i$. So
$\{\ket{\psi_0^{(k)}},\ket{\psi_1^{(k)}}\}_{k=1\cdots r}$ are
mutually orthogonal although $\ket{\psi_0^{(k)}}$ and
$\ket{\psi_1^{(k)}}$ may not. Let $P_k$ be the projector onto ${\rm
span}\{\ket{\psi_0^{(k)}},\ket{\psi_1^{(k)}}\}$ for each $k=1\cdots
r$. Let $P_0=P-\sum_{k=1}^r\op{\psi_0^{(k)}}{\psi_0^{(k)}}$ and
$P_{r+1}=Q-\sum_{k=1}^r\op{\psi_1^{(k)}}{\psi_1^{(k)}}$. Here both
$P_0$ and $P_{r+1}$ may be vanishing. One can readily verify that
$\{P_0,P_1,\cdots, P_r, P_{r+1}\}$ forms a complete projective
measurement on ${\rm supp}(P+Q)$. Applying this measurement to
$\rho_0$ and $\rho_1$. If the outcome is $0$ or $r+1$ then the
original system is in state $\rho_0$ or $\rho_1$, respectively. We
can directly prepare a target state as $\psi_0$ or $\psi_1$.
Otherwise the outcome is $1\leq k\leq r$ and the post-measurement
state should be $\ket{\psi_0^{(k)}}$ or $\ket{\psi_1^{(k)}}$,
depending on the original state is $\rho_0$ or $\rho_1$. Note that
$|\ip{\psi_0^{(k)}}{\psi_1^{(k)}}|\leq \f(\rho_0,\rho_1)\leq
|\ip{\psi_0}{\psi_1}|$. By Ref. \cite{Uhl85} again  we can further
transform $\{\ket{\psi_0^{(k)}},\ket{\psi_1^{(k)}}\}$ into
$\{\ket{\psi_0},\ket{\psi_1}\}$. With that we complete the proof of
the lemma.\hfill $\blacksquare$

{\bf Proof of $W_q(I_d\otimes A)=W_q(A)$} We will introduce a more
manageable representation of $W_q(A)$ first. A relevant notion is
the David-Wielandt shell of $A$, which is the joint numerical range
of $A$ and $A^\dagger A$:
\begin{equation}\label{DW}
DW(A)=\{(\braket{\psi}{A}{\psi},\braket{\psi}{A^\dagger
A}{\psi}):\ip{\psi}{\psi}=1)\}.
\end{equation}
It was known that $DW(A)$ is always convex for $d\geq 3$. For $d=2$,
$DW(A)$ is convex if and only if $A$ is normal. For the case that
$A\in\b(\h_2)$ is not normal, $DW(A)$ is an ellipsoid without
interior in $\mathcal{R}^3$, which is obviously not convex. We can
introduce a scalar function on $W(A)$ as follows:
$$h_A(z)=\max\{t: (z,t)\in DW(A)\}.$$
By the Cauchy-Schwartz inequality, we have $h_A(z)\geq |z|^2$ for
any $z\in W(A)$. More importantly, $h_A$ is a concave function (and
thus continuous) on $W(A)$ for any $A\in \b(\h_2)$, as a consequence
of the convexity of $DW(A)$ for $d>3$ or the geometric observation
on the shape of $DW(A)$ for $d=2$. The significance of $h_A$ is
justified in the following representation of $W_q(A)$ essentially
due to Tsing \cite{Tsi84}:
$$
W_q(A)=\{qz+\bar{q}w\sqrt{h_A(z)-|z|^2}: z\in W(A), |w|\leq 1\},
$$
where $\bar{q}=\sqrt{1-q^2}$. Hence the convexity of $W_q(A)$ for
$0\leq q<1$ follows from the convexity of $W(A)$ and the concaveness
of $h_A$.

From the above representation, it is clear that if $DW(A)=DW(B)$,
then $W_q(A)=W_q(B)$ for any matrices $A$ and $B$, and $0\leq q\leq
1$. However, this requirement can be relaxed. To see this, introduce
the {\it upper boundary} of $DW(A)$, say $\partial DW(A)$, as the
set of $(z,t)\in DW(A)$ such that $(z,t')\not\in DW(A)$ for any
$t'>t$. One can readily see this is exactly the set of
$\{(z,h_A(z)):z\in W(A)\}$. Then it was shown in Ref. \cite{LN98}
that $W_q(A)=W_q(B)$ for any $0\leq q\leq 1$, if and only if
$\partial DW(A)=\partial DW(B)$.

Now let us apply this result to the case of $A$ and $B=I_d\otimes
A$. Our first observation is the following
$$DW(A)={\rm Conv}\{z: z\in DW(A)\},$$
which can be verified directly from the definition of $DW(A)$. Thus
$DW(I_d\otimes A)=DW(A)$ follows immediately if $d\geq 3$ or $d=2$
and $A$ is normal, as $DW(A)$ is convex in this case. For $d=2$ and
$A$ is not normal, we know that $DW(A)$ is an ellipsoid without
interior (by a direct calculation, or see Ref. \cite{LN98}).
Combining this observation with $DW(I_2\otimes A)={\rm
Conv}(DW(A))$, we know $DW(I_2\otimes A)$ is a solid ellipsoid with
$DW(A)$ as its surface. It is clear that $\partial DW(I_2\otimes
A)=\partial DW(A)$ in this case. \hfill $\blacksquare$

\begin{thebibliography}{99}
\bibitem{A-isometry} We can show that a linear operator $A\in
\b(\h_d)$ can be write into the form $A=U_0^\dagger U_1$  for some
isometries $U_0$ and $U_1$ from $\b(\h_d)$ to $\b(\h_{d'})$ iff
$A^\dagger A\leq I_d$.
\end{thebibliography}
\end{document}